# Macropixel-Constrained Collocated Positions and Neighbors Motion Search for Plenoptic 2.0 Video Coding

Vinh Van Duong, Thuc Nguyen Huu, Jonghoon Yim, and Byeungwoo Jeon

***Abstract*—Compared to the unfocused plenoptic 1.0 camera, the newly introduced focused plenoptic 2.0 camera achieves significantly higher spatial resolution of captured light field videos by efficient light field sampling. In order to leverage novel insights on the motion vector distribution in light field videos captured by plenoptic 2.0 cameras, we propose a fast motion estimation methodology for encoding plenoptic 2.0 videos. To further enhance its efficiency, the proposed fast search is optimized by incorporating both macropixel-level and pixel-level search strategies, substantially reducing the encoding time complexity. Experiment results validate the effectiveness of the proposed method, demonstrating significant bitrate savings and computational cost reductions across diverse coding scenarios.**

***Index Terms*— Light field, plenoptic camera, video coding, fast motion estimation, macropixel image.**

## I. Introduction

Light field (LF) imaging captures both the directions and intensities of light rays from scene to an observer using a plenoptic camera. As shown in Fig. 1(a), a conventional camera only captures the average intensity of light rays hitting camera sensors while losing the direction information of light rays. In contrast, a plenoptic camera in which a microlens array (MLA) is placed in front of the image sensor plane can record not only the intensity but also the direction of the light rays. A group of pixels in the camera sensor collected by a microlens is referred to as a macropixel (or micro-image), and a collection of macropixels (see Fig. 1(b)) captured by an MLA is called lenslet image or macropixel image. With this unique optical structure, images or videos captured by a plenoptic camera provide much richer and more usable information for various applications such as depth estimation, point cloud data processing, or saliency detection [1]-[5].

Vinh Van Duong, Thuc Nguyen Huu, Jonghoon Yim, and Byeungwoo Jeon (*corresponding author,* bjeon@skku.edu) are with the Digital Media Laboratory (DMLAB), Department of Electrical and Computer Engineering, Sungkyunkwan University, 2066 Seoburo, Jangan-gu, Suwon, 16410, KOREA.

This work was supported in part by Basic Science Research Program (RS-2023-00208453) through the National Research Foundation of Korea (NRF) and by the ICT Creative Consilience Program (IITP-2025-2020- II201821) supervised by the IITP (Institute for Information & communications Technology Planning & Evaluation), both funded by the Ministry of Science and ICT, Korea.

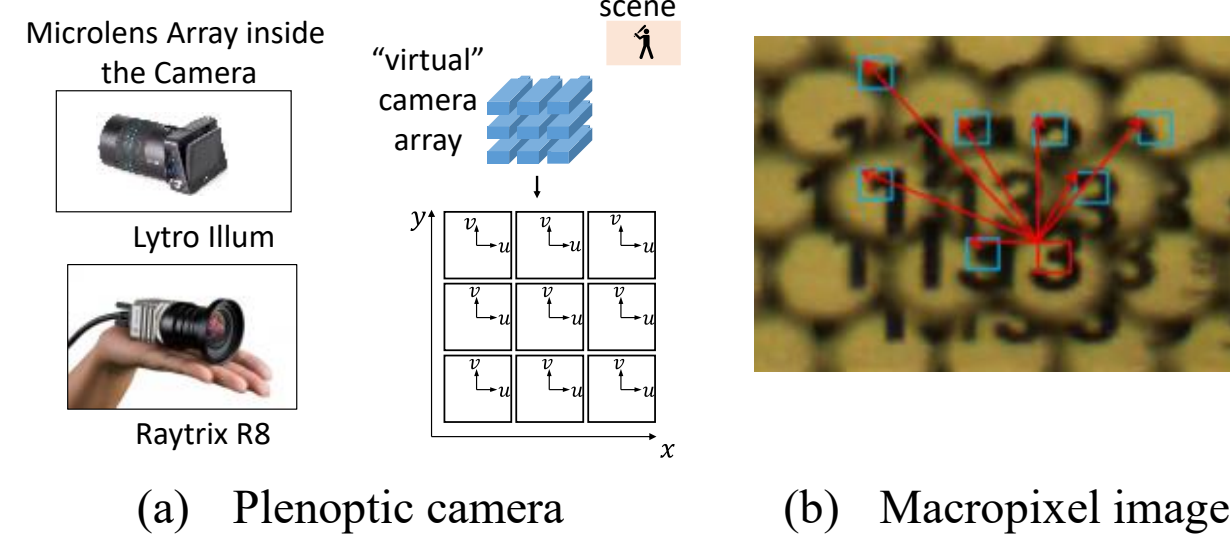

**Fig. 1.** Illustration of plenoptic camera and its captured image.

Because the intensity distribution of a plenoptic image or video is markedly different from that of conventional images or videos, appropriate care should be paid to reduce spatial and temporal redundancies and to maintain fidelity in plenoptic imagery. While a comprehensive survey of coding solutions for plenoptic images exists [7]-[8], plenoptic video coding has become a relatively new research topic, receiving significant attention recently from those working for MPEG-Immersive encoding [9]-[10]. Succeeding the well-known plenoptic 1.0 camera (e.g., Lytro Illum camera) whose commercial availability on the market no longer exists, the research community is focusing on plenoptic 2.0 cameras (e.g., Raytrix camera) which can provide more flexible control of the trade-off between spatial and angular resolutions. There is much redundancy among macropixels in the same picture and also over different pictures. Taking advantages of the distinct intensity distribution of plenoptic 2.0 images, several new intra-coding tools [11]-[14] are already designed for encoding plenoptic 2.0 images. However, inter-coding tools including fast motion estimation (ME) for plenoptic 2.0 video have yet to be investigated more. The distinct optical design of the plenoptic 2.0 camera, compared to the plenoptic 1.0 camera, accounts for the more complex motion behavior observed in plenoptic 2.0 videos. This different motion characteristic highlights the limitations of existing methods developed for plenoptic 1.0 in accurately estimating motion vectors for plenoptic 2.0.

The plenoptic image is captured by 2D array of microlens each of which can be modeled by a pinhole camera as shown in

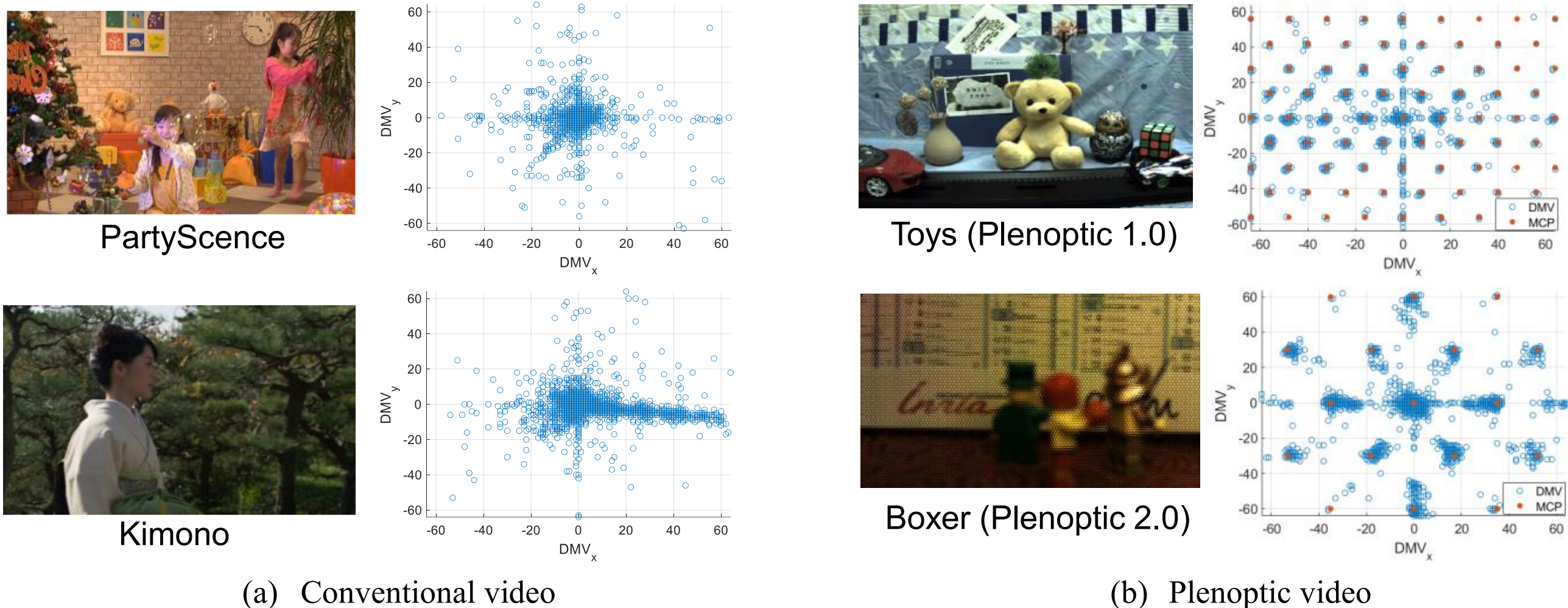

**Fig. 2**. Illustration of typical motion vector distribution of conventional video and plenoptic video. The blue circle corresponds to the difference motion vector (DMV) between a given predicted MV (which is the center of search window) and optimal MV estimated by Full Search. The red dots show the macropixel-constrained collocated positions (MCPs) in plenoptic video corresponding to the given predicted MV. (a) The optimal motion vectors have high population around the given predicted MV. (b) The optimal motion vectors are mostly populated around the MCPs corresponding to the predicted MV. Compared to Plenoptic 1.0 video, Plenoptic 2.0 video exhibits notably larger motion deviations from the MCPs.

Fig. 1(a) where $(x,y)$ indicates the location of each microlens (that is, spatial sampling coordinate), and $(u,v)$ indicates a location inside each microlens (that is, angular sampling coordinate). Due to the special optical structure of plenoptic camera employing lenslet array, one can define a set of pixel positions, called macropixel-constrained collocated positions (MCPs) [15], which share the same $(u,v)$ position in macropixels.

Through statistical analysis of motion distribution, as illustrated in Fig. 2, we observe that the optimal motion vectors (MVs) in plenoptic 2.0 videos, similar to those in plenoptic 1.0 videos, are predominantly concentrated around MCPs corresponding to a center position of search window (that is, predicted MV). However, the deviations of optimal motion vectors from the MCPs are considerably larger in plenoptic 2.0 videos compared to the plenoptic 1.0 cases. Building on this observation, we propose a joint motion search over MCPs and their neighbors as well to effectively address the large motion vector deviations in plenoptic 2.0. It achieves a substantial improvement over established techniques, such as MFME [15] and MTSS [16], by facilitating capture of more precise motion vectors in plenoptic 2.0 videos. This method, while relatively simple, proves to be highly effective since it is grounded in our comprehensive analysis and deep understanding of the motion characteristics specific to plenoptic 2.0 videos. Preliminary results of this paper were reported in the MPEG-Immersion Video activity [17]-[20] and a conference paper [21]. In short, the contributions of this paper are summarized as follows:

- We identify the primary differences in motion vector distributions among conventional videos, plenoptic 1.0 videos, and plenoptic 2.0 videos through a statistical analysis on real captured datasets.
- We propose a novel fast motion estimation method for encoding plenoptic 2.0 videos. It does a joint search over MCPs and their neighbors to effectively address the large motion deviations from MCPs in plenoptic 2.0 video.
- To reduce computational time complexity of motion estimation in a large window search, we propose a macropixel-level diamond search pattern (MLDSP) following the center–biased distribution of motion vector at macropixel-resolution. Additionally, we introduce a fast MCP neighbors search, restricted to the top K-MCP locations with the lowest distortion costs.
- Furthermore, it can be easily generalized to accommodate the pre-processing of plenoptic 2.0 videos, where the pre-processing steps of cropping and alignment are applied to microlens images for codec-agnostic plenoptic video coding [31].

The remainder of this paper is organized as follows. Section II provides an overview of the structures of plenoptic 1.0 and 2.0 cameras, along with their individual motion distribution characteristics. Sections III and IV introduce the proposed fast motion estimation method for plenoptic 2.0 video coding and evaluate its performance in various captured plenoptic 2.0 videos. The conclusion is presented in Section V.

## II. Background Works and Motivations

This section presents background information on the optical structures of plenoptic cameras. Additionally, it offers a comparative statistical analysis of the motion vector distributions in conventional and plenoptic videos to highlight their key differences.

### A. *Structure and Characteristics of Plenoptic Cameras*

Due to the notable differences in the optical designs of plenoptic 1.0 and 2.0 cameras, analyzing pixel intensity distributions provides essential insights for developing efficient and customized motion search techniques specific to each camera type. The plenoptic 2.0 video coding which is a relatively new topic recently gaining attention from the MPEG-Immersion Video (MIV) activity [9][10], requires further investigation to develop effective compression methods tailored to various acquisition devices. Unlike the plenoptic 1.0 camera, the plenoptic 2.0 camera utilizes a microlens array focused on the image plane of the main lens, as depicted in Fig. 3. Each microlens captures a portion of the image projected by the main lens. The sensor is effectively positioned farther from the main lens, allowing the image to form at a distance in front of the microlens array. This microlens array functions as a collection of individual cameras, reimaging specific sections of the main image onto the sensor. Each microlens forms a relay imaging system with the main camera lens. Its position satisfies the lens equation, $1/a + 1/b = 1/f_m$, where $a$ is the distance between the MLA and the image plane of the main lens, $b$ is the distance between the MLA and camera sensor, and $f_m$ is the focal length of the microlens. Using the geometric distance, the distance between the pixel's response on the sensor from the same object point, $s$, can be calculated using $d$ (the diameter of each microlens) which varies depending on the plenoptic camera model.

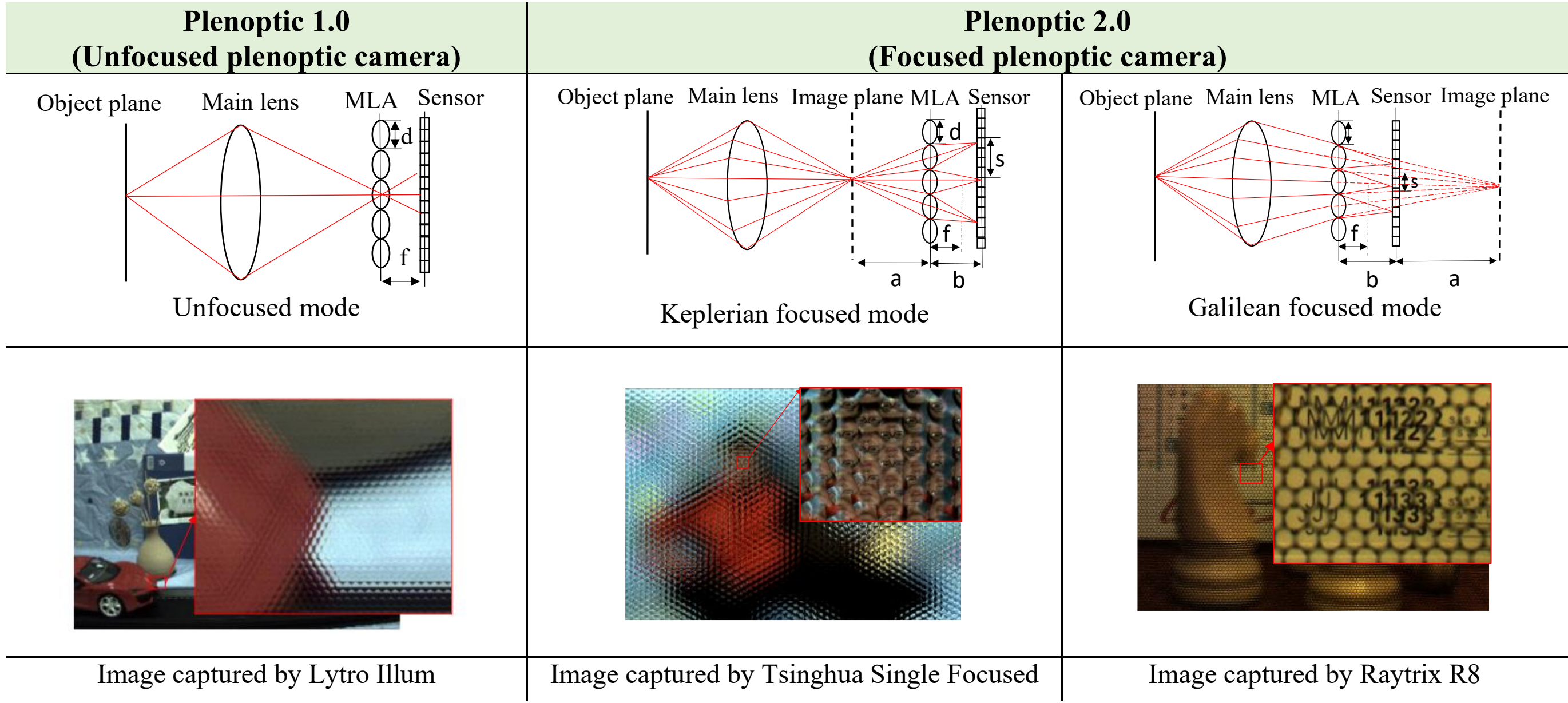

**Fig. 3**. The optical structures and real macropixel images captured by plenoptic 1.0 and 2.0 cameras. The image patterns captured by plenoptic 2.0 cameras are spreading over several neighboring microlens compared to images captured by plenoptic 1.0.

Comparison of the optical designs of plenoptic 1.0 and 2.0 cameras reveals several key distinctions. In a plenoptic 2.0 camera, the relay imaging system ensures that the spatial-angular resolution is not constrained by the number of microlenses. This allows different perspectives for a given spatial point to be captured by different microlenses, enabling significantly higher spatial resolution compared to a plenoptic 1.0 camera. Specifically, rendering with plenoptic 2.0 camera data involves integrating across microlens images unlike the plenoptic 1.0 camera where rendering is performed within individual microlens images. Consequently, the plenoptic 2.0 camera achieves much higher spatial resolution for light field (LF) images than the plenoptic 1.0 camera.

### *B. Statistical Analysis and Motivation*

To analyze the differences in motion characteristics between conventional and plenoptic videos, the VVC reference software VTM-11.0 [28] was utilized for encoding. For conventional videos, we employed several standard video sequences whose properties are detailed in the Common Test Conditions [27]. For the plenoptic video test sequences, we utilized video sequences captured by three different types of plenoptic cameras as in Fig. 3 [29]. The first 30 frames of each test sequence were encoded using VVC reference software VTM-11.0. To determine the optimal motion vectors, the full search was performed with a search area of 64x64. The main observations derived from this analysis are detailed below.

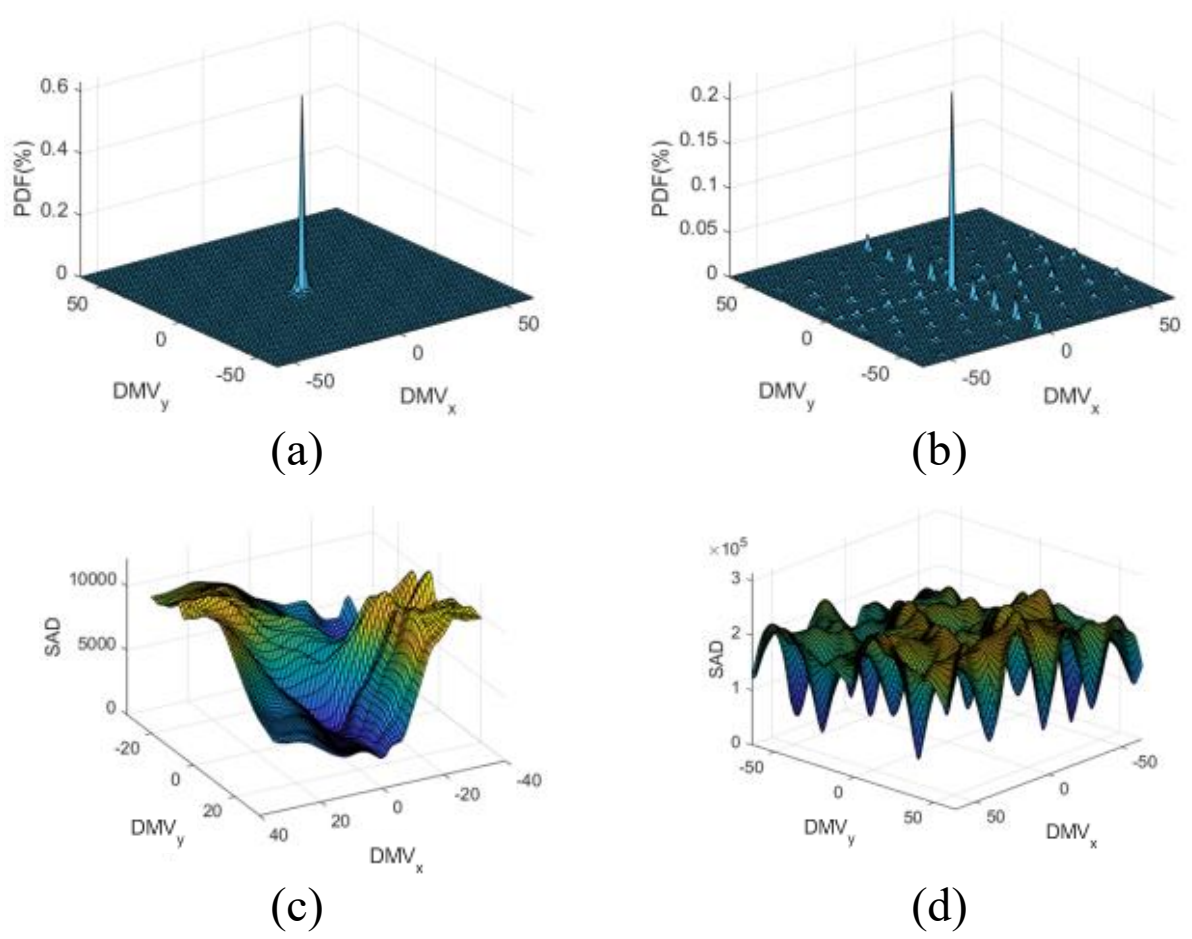

**Fig. 4.** Illustration of typical MV distributions and matching error surfaces of conventional (left side) and plenoptic (right side) videos: the probability density function (PDF) of MV (first row), and the matching error surface in terms of sum of absolute differences (SAD) (second row).

**For plenoptic video, the center-biased assumption of optimal motion vectors in conventional video is still valid but around MCPs**: Comparison of Fig. 4(a) and Fig. 4(b) reveals that the motion vectors of plenoptic video are predominantly concentrated around the center of the search window (that is, around the predicted MV) as in conventional video. However, unlike the conventional video, Fig. 4(b) further reveals that the concentration also exists around MCPs corresponding to the center of search window. This characteristic distinct from conventional video arises from the microlens array elements. Since current frame and its reference frame(s) are highly similar to each other in many cases, motion vectors are much distributed around center of search window (which is the location of predicted MV). However, since each microlens in plenopic camera captures object motion at slightly difference perfective, motion vectors are located also around the MCPs. We refer to this as macropixel-level center-biasedness of MVs in plenoptic video.

**The MVs are more dispersed around the MCPs in plenoptic video 2.0 than in plenoptic 1.0.** We already noted in Fig 2(b) that due to the different optical configurations of plenoptic 1.0 and 2.0 cameras, the optimal motion vectors in plenoptic 2.0 videos deviate more from the MCPs than in plenoptic 1.0. The deviation depends on many factors, and one of them is the different depths of object in the image. Particularly in plenoptic 2,0, not all lenslets have the same focal length, therefore each macropixel does not necessarily capture object as the same size. Depending on the magnitude of MV, optimal MV can deviate more from the MCPs since each microlens in case of plenoptic 2.0 camera records part of an object and creates the defocused macropixel images in the camera sensor as shown in Fig. 3. The defocused macropixel images are spread over several neighboring macropixels. More importantly, the degree of spread is not constant but varies depending on the specific characteristics of each video sequence, such as the disparity range, the depth of field, and the motion of objects. This discrepancy effectively explains the ineffectiveness of the MFME [15] and MTSS [16] methods when applied as fast search processes for encoding plenoptic 2.0 video. Specifically, restricting the search process only to MCPs (depicted as red circles in Fig. 2), as implemented in MFME [15] and MTSS [16], does not guarantee successful identification of an optimal motion vector. This is due to the fact that fewer optimal motion vectors align precisely with MCPs in plenoptic 2.0 videos compared to those in plenoptic 1.0 cases. The substantial motion deviation from MCPs is the distinguishing motion characteristic of plenoptic 2.0 video from the plenoptic 1.0 counterpart.

**The assumption of a uni-modal error surface is no longer valid for plenoptic videos**: In Fig. 4, the sum of absolute difference (SAD) values computed in the block-matching process are shown to highlight their differences between conventional and plenoptic videos. Notably, Fig. 4(d) demonstrates that the assumption of a uni-modal error (i.e., SAD) surface depicted in Fig. 4(c) is no longer valid for plenoptic videos. Consequently, the Test Zone Search (TZS) method [32], initially developed based on the uni-modal assumption, exhibits significantly degraded performance when applied to plenoptic videos. This drawback motivated us to design a novel fast motion search pattern for plenoptic video (named as MLDSP(Macropixel-Level Diamond Search Pattern (MLDSP)).

## III. Proposed Method for Plenoptic 2.0 Videos

This section introduces the proposed motion vector search method for plenoptic 2.0 video, incorporating a newly designed search pattern, MLDSP at each stage as illustrated in Fig. 5. It further demonstrates how the proposed method estimates MV faster than the state-of-the-arts (SOTA) methods like MFME [15] and MTSS [16].

### *A. Search Locations Addressing the Center-biased Assumption at the Macropixel Level*

**Step 1**: **Define MCPs corresponding to a given predicted MV inside a given search window:** To accelerate the motion estimation process, we search MVs only at the selected MCP search candidate positions defined within a specified search window. This approach reduces computational complexity by narrowing the search space for MV, compared to the exhaustive pixel-by-pixel exploration in the Full Search method. Depending on how the microlenses are distributed in MLA, there can be either horizontal or vertical MLA orientation [29]. For simplicity, we describe the process of defining MCP candidates in the horizontal MLA orientation since a similar method can be straightforwardly applied for vertical MLA orientation. For the horizontal MLA orientation in Fig. 6, MCPs are spaced apart vertically by $h = \sqrt{3}\,d/2$ and horizontally by $w = d$. Here, $d$ refers to the microlens diameter. The MCP search candidates inside a given search window with a size ($2W$) can be determined based on Eq. (1) where $(x, y)$ represents the coordinate of an MCP candidate within the search window.

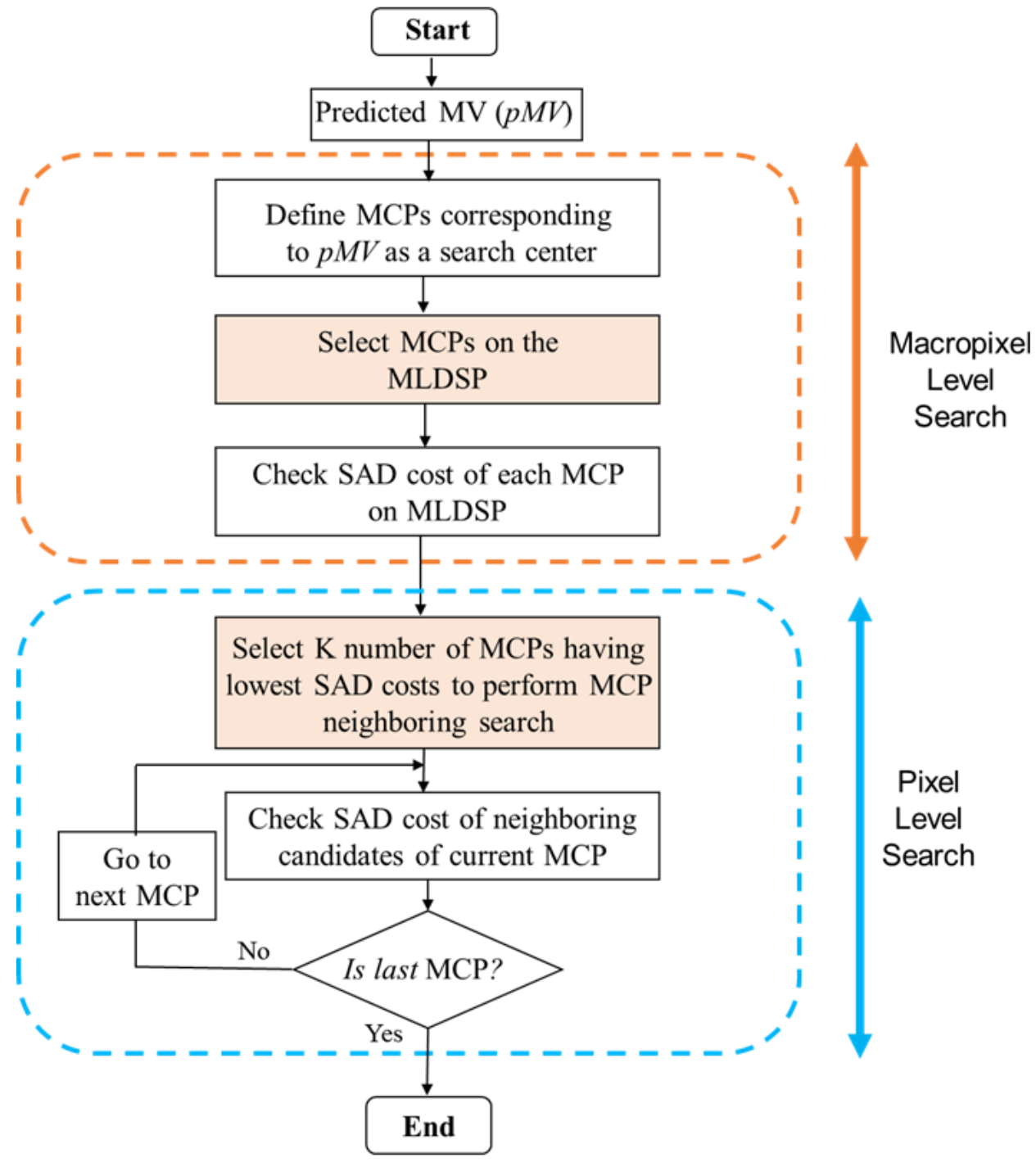

**Fig. 5.** An illustration of our proposed method. The fast search decision in each search level is highlighted in orange.

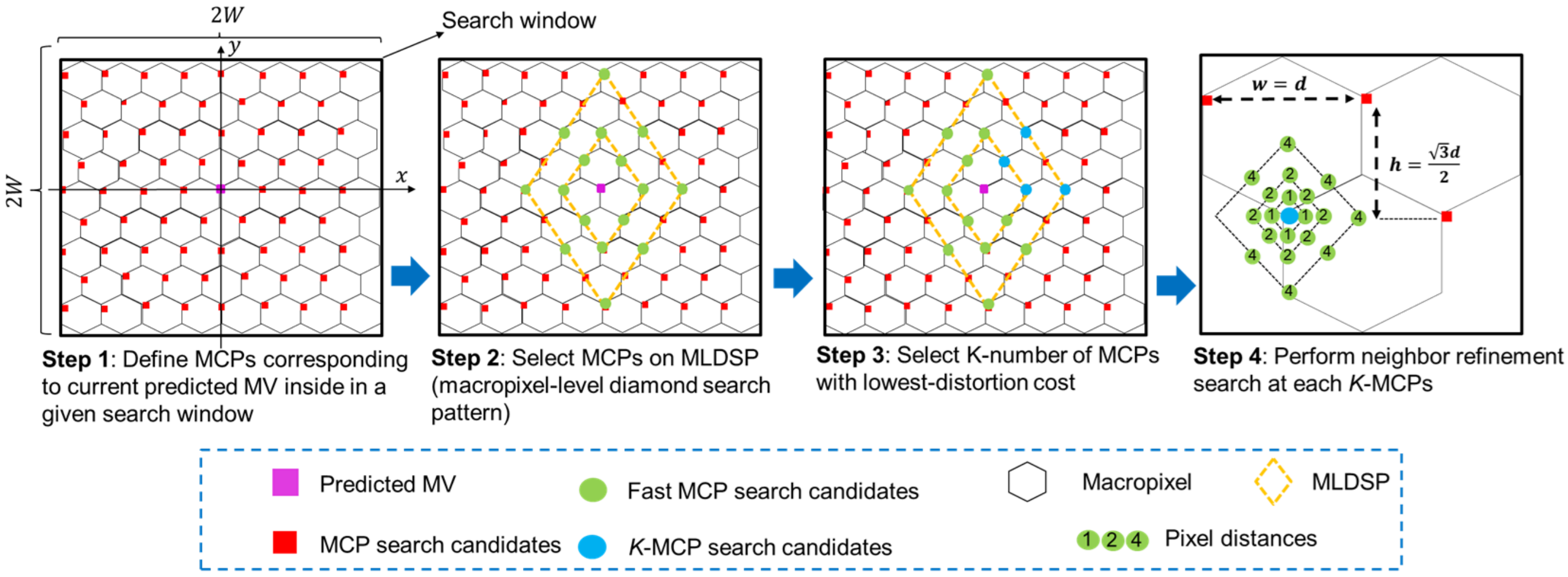

**Fig. 6.** Overview of our proposed fast motion estimation for plenoptic 2.0 video which incorporates new search patterns MLDSP tailored to address the motion characteristics of at both macropixel and pixel search levels.

$$|x| \le W,\ |y| \le W \text{ such that}$$

$$x = \begin{cases}(p + 0.5)\cdot w, & q \in O \\ p \cdot w, & q \in E\end{cases}, \quad y = q \cdot h \tag{1}$$

$$with\ \{p \in \mathbb{Z} \mid |p| \le W/w\}, \{q \in \mathbb{Z} \mid |q| \le W/h\}$$

$O$ and $E$ respectively represent set of odd and even integers, and $\mathbb{Z}$ is a set of integers. The total number of MCP candidates inside the given search window can be computed as follows where $\lfloor . \rfloor$ is the floor operation.

$$\psi_1 = \begin{cases}(m+1)\cdot(2n+1) + 2m\left(n + \dfrac{1}{2}\right), & m \in E \\ m\cdot(2n+1) + 2(m+1)\left(n + \dfrac{1}{2}\right), & m \in O\end{cases} \tag{2}$$

where $m = \left\lfloor \frac{W}{h} \right\rfloor, n = \left\lfloor \frac{W}{w} \right\rfloor$

**Step 2: Select MCPs on macropixel-level diamond search pattern**: In Step 1, MCP candidates corresponding to a given PMV are identified in order to estimate motion vectors which are primarily concentrated at MCPs. This approach follows the straightforward methodology already established in previous works [15][16]. In this study, however, it is observed that the motion vector distribution demonstrates a center-biased tendency at the macropixel level. To effectively avoid from evaluating all MCPs defined in Eq. (1), thereby significantly accelerating MV estimation while keeping the high accuracy, we introduce a novel "macropixel-level diamond search pattern (MLDSP)" and further constrain MCP candidates within the MLDSP. The MLDSP offers distinct advantages particularly for large search windows by significantly reducing the number of candidate MCPs for block matching compared to those defined in Step 1. The MLDSP is designed as follows. An MV can be treated as a pair of discrete random variables $X$ and $Y$ for its horizontal ($X$) and vertical ($Y$) components and the joint cumulative distribution function (CDF) of an MV inside a given search window is defined as:

$$F_{XY}(x, y) = Prob(X \le x, Y \le y), \tag{3}$$

Note that the MV is estimated within the search window, therefore $|x| \le W,\ |y| \le W$, and the joint CDF satisfies $0 \le F_{XY}(x, y) \le 1$. Due to the discrete nature of the random variables $X$ and $Y$, the joint CDF $F_{XY}(x, y)$ is a step function that can jump only at the value that the random variables $X$ and $Y$ can take. For instance, $F_{XY}(x, y)$ jumps at $(x_k, y_k)$ while staying flat between $(x_k, y_k)$ and $(x_{k+1}, y_{k+1})$ where $\varepsilon_k = x_{k+1} - x_k = y_{k+1} - y_k$. Therefore, one can write:

$$F_{XY}(x, y) = F_{XY}(x_k, y_k), \quad with\ x_k \le x < x_{k+1}\ and\ y_k \le y < y_{k+1} \tag{4}$$

The joint CDF of conventional video shows the center-biased distribution nature as in Figure 7(a), and this property has been successfully exploited by allocating more searching points closer to the center and fewer as the distance from the center increases in the fine-to-coarse search strategy in MV estimation. Furthermore the search range is expanded non-linearly to address the non-linear distribution of the joint CDF [24][32]. For example, TZS exploits this property by expanding the diamond search pattern in non-linear step sizes with powers of 2 as shown in Fig. 7(b) [32] since the joint CDF of MV in conventional video jumps at the steps of $\varepsilon_k = 2^k$ where $\{k \in \mathbb{Z} \mid 0 \le k \le \log_2 W\}$. Therefore, the first diamond loop is defined with $\varepsilon_k = 1$ (i.e., $k = 0$), and TZS checks only the four nearest search candidates around the given PMV (center of window search). Subsequently, the diamond search pattern is expanded with the step size increased exponentially ($\varepsilon_k = 2^k$), and it has eight-points to search. The center of diamond remains the same to avoid the local minima problem [32]. When the step

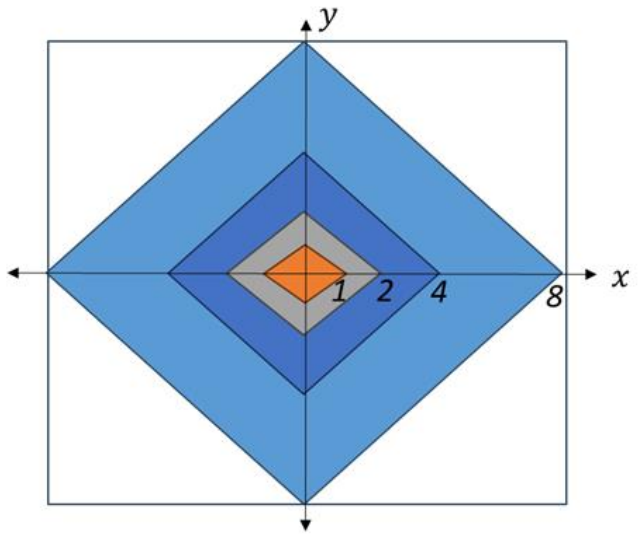

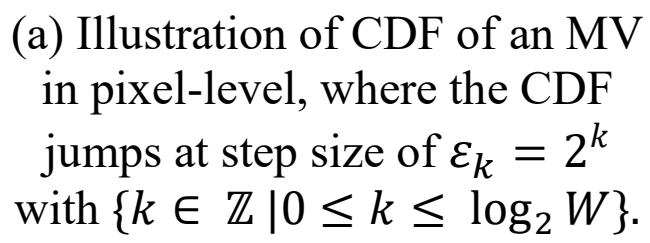

(a) Illustration of CDF of an MV in pixel-level, where the CDF jumps at step size of $\varepsilon_k = 2^k$ with $\{k \in \mathbb{Z} \,|\, 0 \leq k \leq \log_2 W\}$.

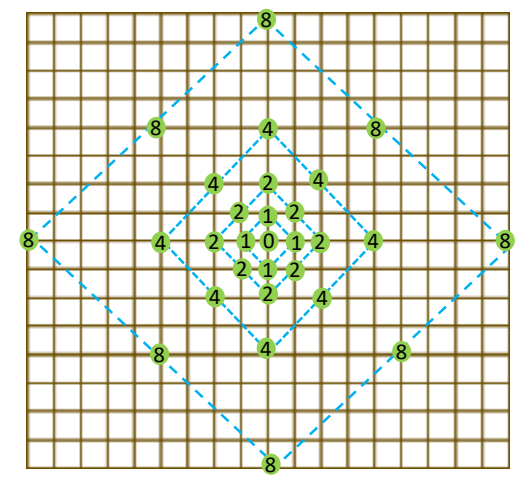

(b) Pixle-level diamond search pattern used in TZS [32]. Note the pixel distance 1→ 2→ 4→8.

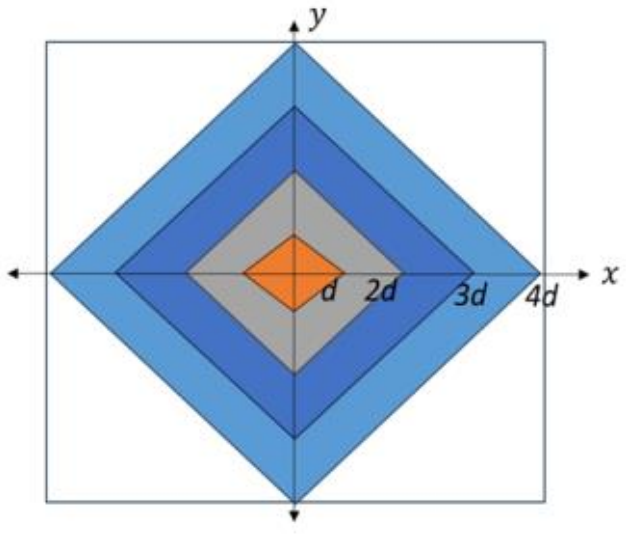

(c) Illustration of CDF of an MV in macropixel-level, where the CDF jumps at step size of $\varepsilon_k = (k+1) \times d$ with $\{k \in \mathbb{Z} \,|\, 0 \leq k \leq \left\lfloor \frac{W}{d} \right\rfloor\}$.

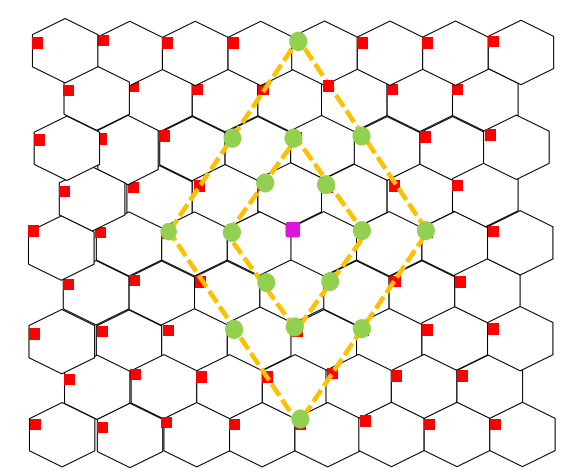

(d) Macropixel-level diamond search pattern (MLDSP) in our method. Note the pixel distance $d$ → $2d$ → $3d$ →$4d$.

**Fig. 7.** Illustration of the cumulative distribution function (CDF) of the estimated MVs following the center-biased assumption at pixel-level resolution and macropixel-level resolution at the top and bottom, respectively.

size reaches the limit of the predefined search range, the MV location having the lowest block matching cost is identified and set as a new center of the diamond in the next refinement step. The same search procedure is applied in the refinement search, where the output of refinement search is the final estimated MV by TZS.

On the other hand, as depicted in Fig. 7(c), the plenoptic video shows slightly different CDF behavior of MV since the jump of CDF occurs at a step whose size increases linearly, that is, $\varepsilon_k = (k+1) \times d$ where $\{k \in \mathbb{Z} \,|\, 0 \leq k \leq \left\lfloor \frac{W}{d} \right\rfloor\}$. It corresponds to the macropixel-level resolution since $d$ refers to the microlens diameter. It fundamentally differs from the CDF jumps at step size $\varepsilon$ with the pixel-level resolution unit (with power of 2) in case of conventional video. Therefore, in plenoptic video, we define a diamond with eight-points with the size of one macropixel at the first loop, while the size of the diamond in the next loop is increasing at a linear scale of macropixel size. The total number of search candidates ($\hat{\psi}_1$) in our MLDSP is calculated as follows:

$$\hat{\psi}_1 = (k+1) \times \eta_1 \quad \text{with } \eta_1 = 8 \text{ and } \left\{k \in \mathbb{Z} \,\middle|\, 0 \leq k \leq \left\lfloor \frac{W}{d} \right\rfloor\right\}, \tag{5}$$

where $\eta_1 = 8$ also corresponds to eight searching directions in the diamond as in pixel-level diamond search pattern in Fig. 7(b) (i.e., two horizontal, two vertical, and two diagonal directions). Note that, since the minimum step size $\varepsilon_0$ in case of plenoptic video is equal to the macropixel resolution ($d$), as shown in Figs. 7(c) and (d), we can define a diamond with eight-directions (eight-points) at the very first searching diamond loop instead of only four-directions (four-points) as in case of conventional video. The MLDSP significantly reduces the number of MCP candidates compared to the predefined MCP search candidates in Eq. (2) (i.e., $\hat{\psi}_1 \ll \psi_1$). For example, $\psi_1$=1307 which is the actual number of MCPs to search in Step 1 calculated by Eq. (2) for the Raytrix R5 camera with $d = 23$ and $W = 384$, is significantly reduced to $\hat{\psi}_1 = \left\lfloor \frac{384}{23} \right\rfloor \times 8 = 128$) by MLDSP.

*B. Search Locations Addressing Large Deviation of Optimal Motion Vectors from MCPs*

**Step 3: Select K-number of MCPs with the lowest block matching cost for refinement search:** A straightforward MV refinement is to search the neighbors around every MCP candidate on MLDSP. In order to avoid much computational demand by doing so, in this paper, we select only the top $K$ number of MCP candidates having the lowest block matching costs. SAD (Sum of absolute difference) is used for the block matching cost. In this paper, we empirically choose $K = 16$ (this choice will be discussed later) which is much smaller than the total number of MCP search candidates in our MLDSP (e.g., $\hat{\psi}_1 = 128$ in the last step).

**Step 4: Perform neighbor MCP refinement search:** Since each microlens in plenoptic camera can be considered as "tiny camera" placed before the camera sensor, the motion behavior within a macropixel is still analogous to that of a conventional single-lens camera. Therefore, in the refinement step, we perform the pixel-level diamond search around each K-MCPs as shown in Fig. 7(b). The number of neighboring candidates at each MCP can be calculated as follows:

$$\psi_2 = \eta_0 + k \times \eta_1 \quad \text{with } \eta_0 = 4, \eta_1 = 8 \text{ with } \left\{k \in \mathbb{Z} \,\middle|\, 0 \leq k \leq \log_2 d\right\}, \tag{6}$$

where $\eta_0 = 4$ corresponds to four-directions in the first diamond loop at pixel-distance of one from the center of search window, and $\eta_1 = 8$ corresponds to eight-directions in the other diamond loops as shown in Fig. 7(b). It should be noted that the search window used in the MCP neighbors search is smaller than that in Step 2 since it is to search the local motion around each MCP. In this step, the search window is set to the microlens diameter ($d$) which is sufficient to account for motion deviation around each MCP. To further avoid redundant search in case of video with slow-motion, an early termination strategy can be applied as in [32].

In summary, the MCP neighbors search operates sequentially at each selected MCP until reaching the final $K$-th MCP. The total number of points for this MCP neighbors search amounts to $K \times \psi_2$. The MCP neighbors search enables one to effectively address large motion deviations occurring around MCPs in plenoptic 2.0 videos. Finally, the total number of search candidates required can be computed as follows:

$$\Omega = \hat{\psi}_1 + K \times \psi_2. \tag{7}$$

TABLE I

SPECIFICATION OF PLENOPTIC 2.0 TEST SEQUENCES [29]

| Camera model | Sequence Name | Lenslet Resolution | Frame rate (fps) | Microlens Diameter (d) | Color Format | Provider |
|---|---|---|---|---|---|---|
| Raytrix R5 | Tunnel | 2048 × 2048 | 30 | 23.30 | YCbCr 4:2:0 | Nagoya University |
| | Origami | | | 23.30 | | |
| | Fujita | | | 23.30 | | |
| | DataLeading | | | 23.20 | | |
| Raytrix R8 | Boxer | 3840 × 2160 | 30 | 35.00 | YCbCr 4:2:0 | INRIA |
| | ChessPieces | | | 35.00 | | |
| | ChessMoving | | | 35.00 | | |
| Single Focused | Boys | 4080 × 3068 | 30 | 72.38 | YCbCr 4:2:0 | Tsinghua University |
| | Experiments | | | 72.38 | | |
| | Cars | | | 70.25 | | |
| | Matryoshka | | | 70.25 | | |

(a) Tunnel (b) Fujita (c) Origami (c) Leading

(e) Boxer (f) ChessPieces (g) ChessMoving

(h) Boys (i) Experimenting (j) Matryoshka (k) Cars

**Fig. 8.** Plenoptic 2.0 video sequences. (a)–(d): Raytrix R5 captured video sequences; (e)–(g): Raytrix R8 captured video sequences; and (h)–(k) Single Focused captured video sequences.

It should be noted that this corresponds to the worst-case scenario in our method (e.g., conditions for early termination in neighbor MCP search is not satisfied). Therefore, the total search candidate in Eq. (6) has remained the same as $\psi_2$.

## IV. EXPERIMENTAL RESULTS AND ANALYSIS

In this section, we evaluate the performance of the proposed method in terms of compression efficiency and computation complexity. Additionally, we present the result of the ablation study.

### *A. Test Conditions*

As shown in Fig. 8, we used eleven plenoptic 2.0 videos from the MPEG-I public dataset [29], which were captured by three kinds of plenoptic 2.0 camera models: Raytrix R5, Raytrix R8, and Single Focused cameras. The specification of each tested sequence is detailed in Table I. The proposed method, along with other competing approaches, was implemented on VVC reference software VTM 11.0 [28], and evaluated under the random access-main (RA-main) configuration following the common test conditions [29] on servers running on the Windows 10 operating system using an ADM Ryzen 9 3900X 12-Core Processor and 32 GB of DDR4 memory. The rate-distortion performance is evaluated in terms of the Bjøntegaard delta bitrate (BDBR) [30], where a negative value denotes better compression performance of the test method compared with the anchor which is the TZS [32]. PSNR (peak signal-to-noise ratio) value used in the BDBR is measured in the Y channel. The encoding time ratio ($ETR$) of a method $m$ is computed for each video sequence as follows:

$$ETR(m) = \frac{T_m}{T_{anchor}} \times 100\ (\%) \tag{8}$$

where $T_m$ and $T_{anchor}$ are respectively the encoding time of a given test method $m$ and the anchor method. The mean of the encoding times of four quantization values are used in the ratio.

To assess the effectiveness of the proposed method, we compare our method with several state-of-the-art fast motion estimation methods for plenoptic videos, including the MFME [15] for plenoptic 1.0 videos and MTSS [16] for plenoptic 2.0 videos. Since the source code of MFME [15] is not available, we re-implemented it in the VVC platform, while the MTSS [16] results were generated using an executable file provided by its authors. To ensure a fair comparison, all methods - FS (full search), MFME [15], MTSS [16], and our method - replace only

TABLE II

PERFORMANCE COMPARISON IN BDBR OF THE PROPOSED AND OTHER METHODS UNDER RA-MAIN CODING CONDITION WITH TZS [32] AS ANCHOR

| | Full Search | MFME[15] | MTSS[16] | **Proposed** |
|---|---|---|---|---|
| Tunnel | -3.52% | -3.43% | -3.07% | -3.34% |
| Fujita | -4.02% | -3.00% | -3.62% | -3.92% |
| Origami | -5.40% | -4.41% | -4.11% | -4.39% |
| Leading | -3.71% | -3.34% | -3.07% | -3.70% |
| Boxer | -2.11% | -1.63% | -1.59% | -1.85% |
| ChessPieces | -4.71% | -4.41% | -4.23% | -4.60% |
| ChessMoving | -4.86% | -3.38% | -4.61% | -4.71% |
| Boys | -0.45% | -0.22% | -0.23% | -0.24% |
| Experiments | -0.87% | -0.50% | -0.64% | -0.62% |
| Cars | -0.02% | 0.01% | -0.02% | -0.03% |
| Matryoshka | -0.81% | 0.27% | -0.39% | -0.07% |
| **Average BDBR** | **-2.77%** | **-2.19%** | **-2.37%** | **-2.50%** |
| **Average ETR** | **1040%** | **99%** | **105%** | **102%** |

TABLE III

ABLATION STUDY ON EFFECT OF DIFFERENT FAST SEARCH CONFIGURATIONS OF OUR METHOD IN RA-MAIN CONDITION WITH TZS [32] AS ANCHOR

| Search level / Case number | Macropixel level search | | Pixel level search | | Average BDBR (%) | Average ETR (%) |
|---|---|---|---|---|---|---|
| | MCP search | Fast MCP search (MLDSP) | Neighbors MCP search | Fast neighbors MCP search (K-MCP) | | |
| **Case I** | O | - | O | - | -2.69% | 112% |
| **Case II** | O | - | - | - | -2.04% | 100% |
| **Case III** | - | O | O | - | -2.59% | 108% |
| **Case IV*** | - | O | - | O | -2.50% | 102% |

*The method adopted in this paper.

the TZS method in the uni-predictive search process. As FS delivered optimal performance, it allows us to evaluate the maximum potential of each method relative to FS.

### B. *Experimental Results*

Table II compares the performance of the proposed method with others under RA-main condition [28] against the anchor of TZS search. From these results, several key conclusions can be drawn. Firstly, our method surpasses the TZS method, yielding an average improvement of 2.50% in terms of bitrate savings in RA-main condition. The TZS method, which is originally designed for conventional videos, exhibits decreased coding performance due to its suboptimal search location and strategy. Notably, our method demonstrates better coding performance compared to the existing methods and closely matches the performance of the FS method. Secondly, our method continues to deliver better performance compared to MFME [15] and MTSS [16]. This advantage arises from the additional search process of our method, which is centered around MCPs and effectively addresses the substantial MV deviations in plenoptic 2.0 videos.

Compared to MFME [15] and MTSS [16], our method yields higher BDBR by approximately 0.31% and 0.13%, respectively. In summary, given the lower coding efficiency of MFME [15] and MTSS [16] compared to FS, these methods may achieve suboptimal performance in the context of encoding plenoptic 2.0 videos when compared to our approach. In terms of time complexity, while our method did not attain the fastest encoding times compared to the existing methods, it consistently demonstrated competitive performance. To be specific, our approach requires approximately 2% and 3% more encoding time in RA-main condition when compared to the TZS [32] and MFME [15] methods, respectively, as shown in last row of Table II. The slight increase in encoding time arises from the neighbors search process of MCP in our method. By employing an innovative fast search approach for MCPs and their neighboring locations, our method achieves a good balance between coding performance and encoding complexity. Compared to MTSS [16], our method is about 3% faster in terms of encoding complexity, while our method still achieves better coding gain demonstrating the effectiveness of our proposed MLDSP and neighbors MCP search processes.

### C. *Ablation Study*

We present the results of the ablation study on the impact of key configurations in our proposed method in Table III. The key conclusions are summarized as follows:

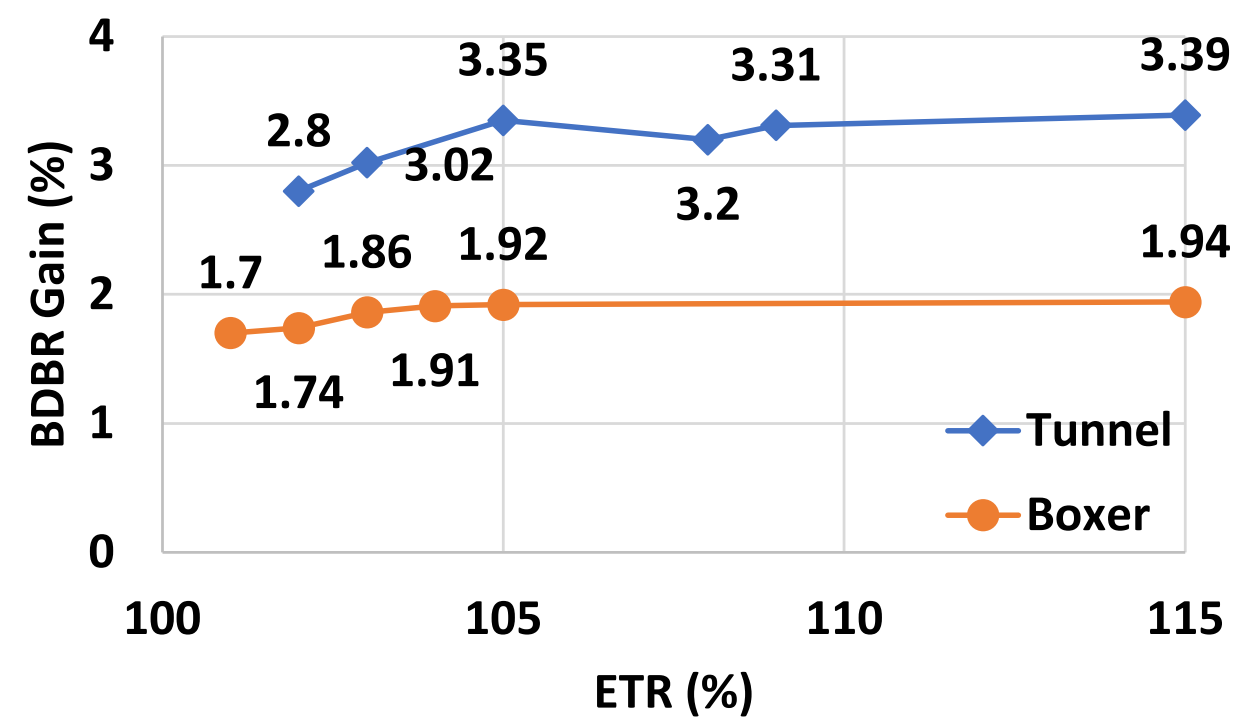

**Fig. 9**. Ablation tests on the parameter $K$ number of neighbors MCP candidates. The graph shows six values of $K \in \{1,2,4,8,16,32\}$ from the left to right for each tested sequence of Tunnel and Boxer. In this paper, we adopted $K = 16$ as the final setup.

**Ablation study on the proposed neighbors search around each MCP (Case I vs. Case II):** In order to highlight the significance of our proposed neighbors search of MCP, we conduct a test by disabling different search locations in our method. Its test result is presented in Table III, where the performance of Case I (only search at the marcopixel-level search locations) exhibits a significant drop of about 0.65% in terms of bitrate savings in comparison to the Case II (search both at the macropixel-level and pixel-level search locations). The reason for this performance decrease is the failure to accurately estimate the optimal motion vectors, particularly in scenarios with substantial motion deviations around each MCP.

**Ablation study on the proposed macropixel-level diamond search pattern approach (MLDSP) (Case II vs. Case III)**: To validate the contribution of the MLDSP, Table III compares the performance of the original MCP and the MLDSP. When MLDSP is employed (Case III), while it suffers BDBR loss of only 0.1% compared to the original MCP search pattern (Case I), encoding time reduction amounts to approximately 4%. It highlights the critical role of the MLDSP in our method.

**Ablation study on the proposed refinement search with *K*-MCP candidates (Case III vs. Case IV):** A new test is conducted to assess performance in the absence of our proposed reduction of neighbors in refinement search. Case IV shows encoding time reduction of 6% compared to Case III. Fig. 9 further illustrates the effect of varying the number $K$. A lager $K$ shows better BDBR but with non-linearly increasing encoding

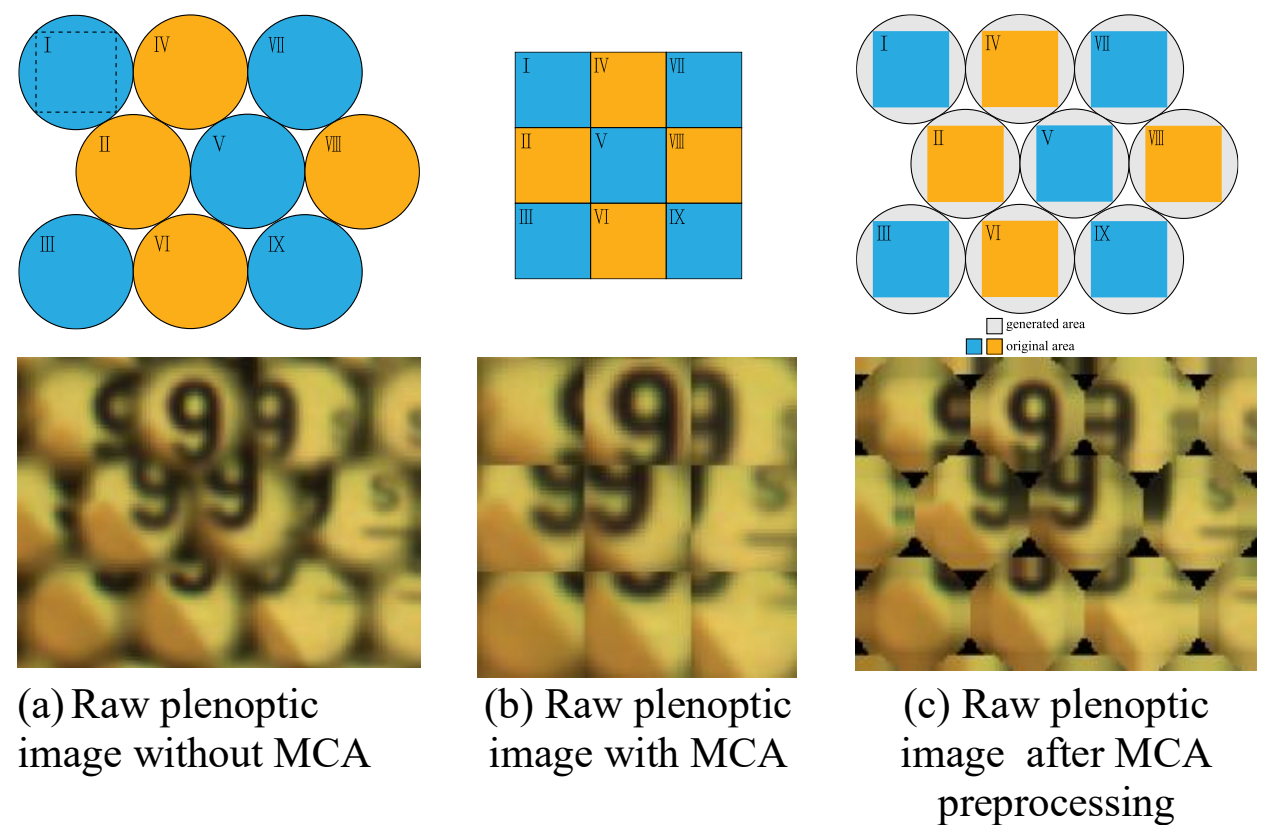

(a) Raw plenoptic image without MCA

(b) Raw plenoptic image with MCA

(c) Raw plenoptic image after MCA preprocessing

**Fig. 10.** Illustration of raw plenoptic image before and after applied microlens cropping and alignment (MCA) method in [31].

time. Considering balance between BDBR performance and encoding complexity, we adopted $K = 16$ in this paper.

### D. *Incorporation with other Pre-processing Tool*

Recently, in MPEG-I [31], the pre-processing of lenslet coding has been studied alongside the original lenslet data format, as illustrated in Fig. 10. In our implementation of expeirment, we apply a Microlens Cropping and Alignment (MCA) [31] module as a preprocessing step to raw plenoptic video data prior to compression. Due to vignetting effects, the pixel regions located near microlens boundaries exhibit significant intensity degradation, often appearing as darkened corners. These regions barely contribute useful information and can be thus identified as redundant. We eliminate these pixels through microlens-aware cropping to reduce input redundancy and improve encoding efficiency. Subsequently, the remaining pixels are resampled and aligned onto a regular Cartesian grid, replacing the native hexagonal sampling pattern of the plenoptic sensor. This alignment is essential for ensuring compatibility with conventional video codecs, which assume regularly sampled input. By applying MCA, we enable a codec-agnostic processing pipeline that bridges the gap between plenoptic data acquisition and standard compression frameworks. To further evaluate our proposed method, we present its coding performance for both the original and pre-processed lenslet data in Table IV. Specifically, we report on the overall performance of our final method under the RA-Main condition. Our proposed method surpasses the TZS anchor, achieving average BDBR gains of -2.50% when encoding raw plenoptic video without MCA and -0.66% with MCA. These coding gains highlight the effectiveness and generalization capability of our proposed approach. This means that our method can be useful not only for encoding the raw plenoptic video but also for the pre-processing ones.

## V. Conclusion

In this paper, we introduced a novel fast MV search method for encoding plenoptic 2.0 videos, leveraging new insights into the motion vector distribution of plenoptic 2.0 content. The proposed method performed a joint search over MCPs and their neighboring regions to effectively address the large motion deviations commonly observed in Plenoptic 2.0 video content. To reduce the computational complexity of ME within a large search window, we introduced a new search method called as MLDSP, which followed the center-biased distribution of motion vectors at the macropixel resolution. Additionally, we implemented a fast MCP neighbor search scheme, restricted to the top K-MCP locations with the lowest distortion costs.

It is worth noting that the proposed method could be also extended to support block vector estimation in the context of intra block copy (IBC). As demonstrated in our previous conference paper [23], the proposed fast search strategy was particularly effective when applied to extended search ranges for IBC. The method achieved notable coding gains while maintaining a favorable trade-off between compression efficiency and computational complexity.

TABLE IV

Overall Performance of Our Proposed Method with The Pre-processing Tool [31] in RA-Main Condition with TZS [32] as anchor.

| Data types / Methods | Apply Microlens cropping and alignment (MCA) [31] | | | |
|---|---|---|---|---|
| | without MCA | | With MCA | |
| | Full Search | **Proposed** | Full Search | **Proposed** |
| Avg. BDBR | -2.77% | -2.50% | -1.03% | -0.66% |
| Avg. ETR | 1040% | 102% | 962% | 102% |